\documentclass[10pt]{article}
\usepackage{emulateapj}
\usepackage{times}
\usepackage{psfig}
\usepackage{natbib}

\newcommand{\be}{\begin{eqnarray}}
\newcommand{\ee}{\end{eqnarray}}

\newcommand{\msun}{\mbox{${\rm M}_\odot$}}
\newcommand{\Msun}{\mbox{${\rm M}_\odot$}}

\newcommand{\trlx}{\mbox{$t_{\rm rlx}$}}

\def\apgt{\ {\raise-.5ex\hbox{$\buildrel>\over\sim$}}\ }
\def\aplt{\ {\raise-.5ex\hbox{$\buildrel<\over\sim$}}\ }
\def\lt{\ {\raise-.5ex\hbox{$\buildrel>$}}\ }
\def\gt{\ {\raise-.5ex\hbox{$\buildrel<$}}\ }

\newcounter{lastnote}

\lefthead{Portegies Zwart}
\righthead{The lost siblings of the Sun}

\begin{document}
\title{The lost siblings of the Sun}

\author{
 Simon F. Portegies Zwart}

\affil{
 	Astronomical Institute `Anton Pannekoek', 
 	University of Amsterdam, Kruislaan 403, Amsterdam, the Netherlands }
\affil{
 	Institute for Computer Science,
	University of Amsterdam, Kruislaan 403, Amsterdam, the Netherlands }
\affil{
 	Sterrewacht Leiden
	University of Leiden, Niels Bohrweg 2, 2333 CA Leiden, the Netherlands
}

\authoremail{E-mail: S.F.PortegiesZwart@uva.nl}

\date{}
\begin{abstract}

The anomalous chemical abundances and the structure of the
Edgewood-Kuiper belt observed in the solar system constrain the
initial mass and radius of the star cluster in which the sun was born
to $M\simeq500$ to $3000$\,\msun\, and $R\simeq 1$ to $3$\,pc. When
the cluster dissolved the siblings of the sun dispersed through the
galaxy, but they remained on a similar orbit around the Galactic
center. Today these stars hide among the field stars, but 10 to 60 of
them are still present within a distance of $\sim 100$\,pc. These
siblings of the sun can be identified by accurate measurements of
their chemical abundances, positions and their velocities. Finding
even a few will strongly constrain the parameters of the parental star
cluster and the location in the Galaxy where we were born.
 
\end{abstract}

\section{Introduction}

It is commonly accepted that stars like the sun are born in clusters
\citep{1993prpl.conf..245L}.  The cluster in which the sun was born is
long gone and the sun's siblings are by now spread over the Galaxy.
The structure of the hot Edgeworth-Kuiper belt objects provide
evidence of this, as this population can be reproduced by a relatively
nearby encounter with another star \citep{2004AJ....128.2564M}.  Such an
encounter is expected to have occurred in the early history of the
Solar system \citep{2008DPS....40.3801M}.  The existence of a well
organized planetary system, however, indicates that the parental
cluster cannot have been very dense as otherwise a nearby passing star
would also have perturbed the orbits of the planets.

Additional evidence for the sun's dynamic history comes from the
discovery of radioactive isotopes and their decay products in the
proto-solar nebula \citep{2004Sci...304.1116H}, which is explained by a
supernova explosion within 1.6\,pc of the infant
sun \citep{2006ApJ...652.1755L}. The combination of arguments enables us
to estimate the mass and size of the star cluster in which the sun was
born. We follow the orbital evolution of these stars through the
Galaxy for the lifetime of the sun and conclude that at least 1\% (10
to 60) of the sun's siblings should still be present within 100\,pc,
and more than 10\% should be within about a kpc along the orbit of the
solar system in the Galactic potential.  With the Gaia
satellite \citep{2001A&A...369..339P} and ground-based searches the
radial velocity and distance to the majority of the lost siblings will
be determined, and the proper motion will be measured. These constrain
the orbit of the proto-solar cluster and enable us to accurately
determine the evolution of the Galactic potential and the birth place
of the sun.

\section{The parental star cluster}

The sun and its eight planets were born about 4.57\,billion years
ago \citep{2002A&A...390.1115B}, probably in a star cluster, and it is
in due time that our location in the Galaxy became so desolate.
Evidence for this dynamic past comes from meteorite fossil records,
where the presence of short-lived radioactive isotopes in primitive
meteorites indicate that the 1.8\,Myr young sun was polluted by a
supernova explosion of a star about 15 to 25 times more massive than
the sun within a distance of 0.02 to 1.6 parsec
 \citep{2006ApJ...652.1755L}.  Such a massive star lives for 6 to
12\,Myr before it sheds the majority of its mass in a supernova
explosion. This massive star must have formed some 4.2 to 10.2\,Myr
earlier than the sun. Such range in the distribution of stellar ages
is also observed in the Orion Nebula where massive stars also tend to
be a few to $\sim 10$\,Myr younger than the low-mass stars
\citep{1999ApJ...525..772P}.

The presence of a massive star close to the infant sun puts
interesting constraints on its birth environment.  Today, such a
massive star is not even present within a distance of 100 parsec.  By
adopting a standard initial mass function \citep{2003ApJ...598.1076K}
about 1 in 400 stars is sufficiently massive to experience a
supernova. Stars of $m = 15$ to 25\,\msun\, are less common and would
require a star cluster of $M \apgt 500$\,\Msun\,
\citep{2004MNRAS.348..187W}.

Massive stars in a cluster tend to sink to the center in a fraction
$\propto 1/m$ of the two-body relaxation time scale ($\trlx$).  If
$\trlx \apgt 300$\,Myr even the most massive stars are unlikely to
have reached the cluster center by the time they explode.  Massive
stars in clusters with smaller \trlx\, tend to populate the central
region by the time they explode in supernovae, polluting the nearby
young stars and proto planetary disks in the process
\citep{1996A&A...314..438W}.  Low mass stars, like the sun, are not
strongly affected by mass segregation, and if such a star happens to
be in the cluster core at the moment one star explodes it is likely to
be still around when a subsequent supernova occurs.

We quantify this by performing a number of $N$-body simulations in
which we initialized star clusters with a mass function and a Plummer
sphere density distribution \citep{1911MNRAS..71..460P} with radius $R$
in parsec to track the number of supernovae that occurred within
1.6\,pc of sun-like main-sequence stars.  It turns out that the number
of supernovae with one and the same star of 0.8 to 1.2\,\msun\, scales
as $N_{\rm nearby SN} \sim 0.003M/R$.

A star cluster is a dynamic environment in which frequent encounters
effectively ionize planetary systems \citep{2007MNRAS.378.1207M}.  The
passage of a star within a distance of $r_{\rm enc} =100$\,
Astronomical Units (AU) can easily destroy the young proto-planetary
nebula, leaving the sun without any material to form a regular
planetary system \citep{1985AJ.....90.1876H,2001Icar..153..416K}. If
the planetary system is already formed such an encounter may induce
high eccentricities and inclinations of their orbits, again resulting
in strong perturbations which eventually leaves the sun with a reduced
planetary system \citep{2006astro.ph.12757S}.  The high eccentricities
and inclinations observed in several extra-solar planetary systems may
be produced by such a relatively nearby encounter. The outer region of
the solar system shows evidence for a relatively nearby encounter but
not as close as 100\,AU \citep{2000ApJ...528..351I}, an encounter at a
distance within $r_{\rm enc} \simeq 10^3$\,AU would be sufficient to
explain the observed hot Kuiper belt objects, which have highly
inclined orbits \citep{1988AJ.....96.1127M,2004Sci...304.1116H}.

\begin{figure}
\psfig{figure=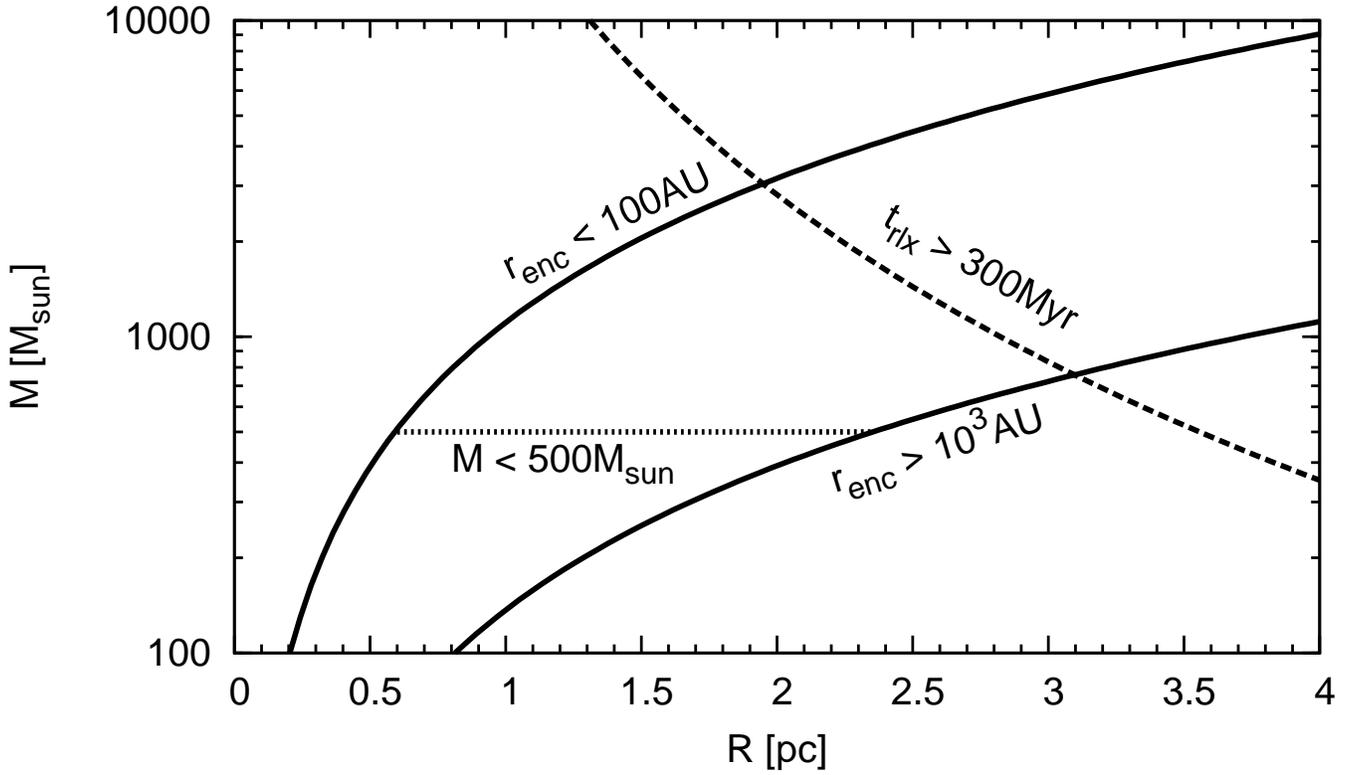,width=\columnwidth}
\caption[]{ 
   Constraints on the mass ($M$ in $\Msun$) and radius ($R$ in pc) of
   the cluster in which the sun was born. The allowed parameter space
   is fenced off by four curves, one for the minimum mass of about
   500\,\msun\, (dotted line), one limits the relaxation time
   ($\trlx$) to less than about 300\,Myr (dashed curve), and two
   limiting the distance between which an encounter is likely to occur
   during the cluster lifetime (solid curves). If an encounter is
   expected outside $r_{\rm enc} = 10^3$\,AU the orbits beyond Pluto
   are unlikely to have been affected, contrary to the observed high
   inclinations in the Kuiper belt \citep{2000ApJ...528..351I}.  An
   encounter within $r_{\rm enc} = 100$\,AU is likely to be
   destructive to the solar system as we know it
   \citep{2006ApJ...641..504A}.  In order to warrant one or more close
   supernovae near the sun before the cluster dissolves we require
   that the relaxation time is smaller than 300\,Myr. The area fenced
   off by the dotted, the dashed and the solid curves is the favorable
   range of parameters for the proto-solar cluster.
\label{Fig:ClusterMass}
}
\end{figure}

The requirement that the sun has not encountered another star at
$r_{\rm enc} \aplt 100$\,AU, but is likely to have experienced an
encounter within some $r_{\rm enc} \aplt 10^3$\,AU constrains the
initial cluster density.  We further require that the cluster survives
long enough to warrant this encounter to occur before it dissolves.
The dissolution time of a star cluster in the Milky Way Galaxy follows
the empirical relation \citep{2005A&A...429..173L} $t_{\rm diss} \simeq
2.3 {\rm Myr}\, M^{0.6}$.  These constraints can be expressed in terms
of cluster radius $R$ (in parsec) and mass $M$ (in \Msun), which we
present in fig.\,\ref{Fig:ClusterMass}. With estimates for the mass
and the size of the parental star cluster we calculate that the
velocity dispersion of the cluster must have been of the order of a
km/s, which is comparable to the velocity dispersion in known young
star clusters \citep{2003ARA&A..41...57L}.

\placefigure{Fig:ClusterMass}

\section{The birthplace of the sun}

The orbit of the solar system in the Galactic potential is almost
circular, more or less in the disk.  The orbital velocity is
considerably higher than the velocity dispersion of the stars in the
parental cluster. The cluster members are therefore unlikely to have
drifted very far from the orbit of the sun around the Galactic center,
but they may be at completely different locations along this orbit.
The preservation of phase space in the dissolution of star clusters is
used to study the merger history of the Galaxy
\citep{1999Natur.402...53H}, but in this case we use the argument to
find the lost siblings of the sun.

At the moment the sun is located at a distance of about 8.5\,kpc from
the Galactic center in the plane of the disk. The velocity is
10.1\,km/s directed towards the Galactic center and 7.5\,km/s away
from the plane. The velocity perpendicular to the line-of-sight of the
Galactic center is $v_{\rm t} = 235.5$ km/s in the rotation direction
of the Galaxy, which is slightly ($\sim 15$km/s, since $v_{\rm c}
\simeq 220$km/s) higher than the local orbital velocity
\citep{1996AstL...22..455K}.

By calculating the orbit of the sun backwards in time, which can be
achieved by inverting the current velocity vector, and integrating the
equations of motion in the potential of the Galaxy for 4.6\,Gyr, we
can calculate the location in the Galaxy where the sun was born. With
the parameters given and using a model for the Galactic potential
\citep{1990ApJ...348..485P} it turns out that the sun has been orbiting
the Galactic center some 27 times since its formation, and was born in
the $\delta$-quadrant at -1.39kpc, 9.34kpc and 25.3pc along the
galactic x, y and z-axis and with a velocity vector of -207, -48.2 and
-6.72km/s (see fig.\,\ref{Fig:TopView}).  This birthplace is $\sim
2.8$\,kpc further out than the estimated distance to the Galactic
center based on the relatively high metalicity of the sun
\citep{1996A&A...314..438W,1997A&A...326..139W}, but part of this
metalicity argument could be explained by the early pollution of a
supernova near the proto-solar nebula.  Part (up to about half) of the
other cluster members are also expected to be polluted by the same
supernova, and their anomalous chemical characteristics can be used for
the identification \citep{2004ApJ...617L.119T}.

It seems likely that the orbits of the sun and it siblings have been
deflected by some small-angle scatterings on their long journey
through Galactic disk. The radial distance over which the sun may have
been migrated by such scatterings can be as large as 2\,kpc
\citep{1987gal..proc..375F}, which interestingy is comparable to the
range in radial distance for the orbit of the sun, as we present in
Fig.\,\ref{Fig:TopView}. The radial velocity induced by such
scatterings affect the orbit of all stars in the parent cluster and
the effect on the number of nearby stars today is rather small. We
quantify this by varying the current tangent velocity of the sun
$v_{\rm t}$ and recompute the orbits of its siblings in the potential
of the Galaxy assumig that the cluster dissolved 4.3\,Gyr ago. The
fraction of stars of the parent cluster that remain within 100\,pc can
then be expressed as a linear relation $f(r < 100{\rm pc}) \simeq
0.032 - 0.014 (v_{\rm t}/v_{\rm c})$, and the fraction of stars within
1kpc becomes $f(r < 1{\rm kpc}) \simeq 0.39 - 0.16 (v_{\rm t}/v_{\rm
  c})$. These relations break down then $v_{\rm t}$ is either so small
that the stars enter the Galactic bulge on their orbit, in which case
the scattering becomes severe, or if $v_{\rm t}$ becomes sufficiently
high that the stars escape the Galaxy.

The uncertainty in the current coordinates of the sun in the Galaxy
will be magnified by computing its orbit back with time, making the
location of birth uncertain.  It turns out, however, that the exact
place and velocity of birth are not crucial for our discussion, as the
stars that were born together with the sun will be around wherever the
sun was born. It is the orbits of those stars for which we integrate
the equations of motion in time through the Galaxy to understand the
dispersion in their position and velocity at a later time.  The
adopted model for the Galactic potential does not include spiral arms
or local perturbations like star clusters and molecular clouds, which
can have an appreciable effect on the calculated trajectories
\citep{1987gal..proc..375F,2005AJ....130..576Q}.  The uncertainties
introduced by the model, however, have a minor effect on the relative
position and velocities as all stars remain relatively close together
in phase space throughout the integration and are therefore affected
by variations in the potential in a similar fashion.

\section{Finding the lost siblings of the sun}

\begin{figure}
\psfig{figure=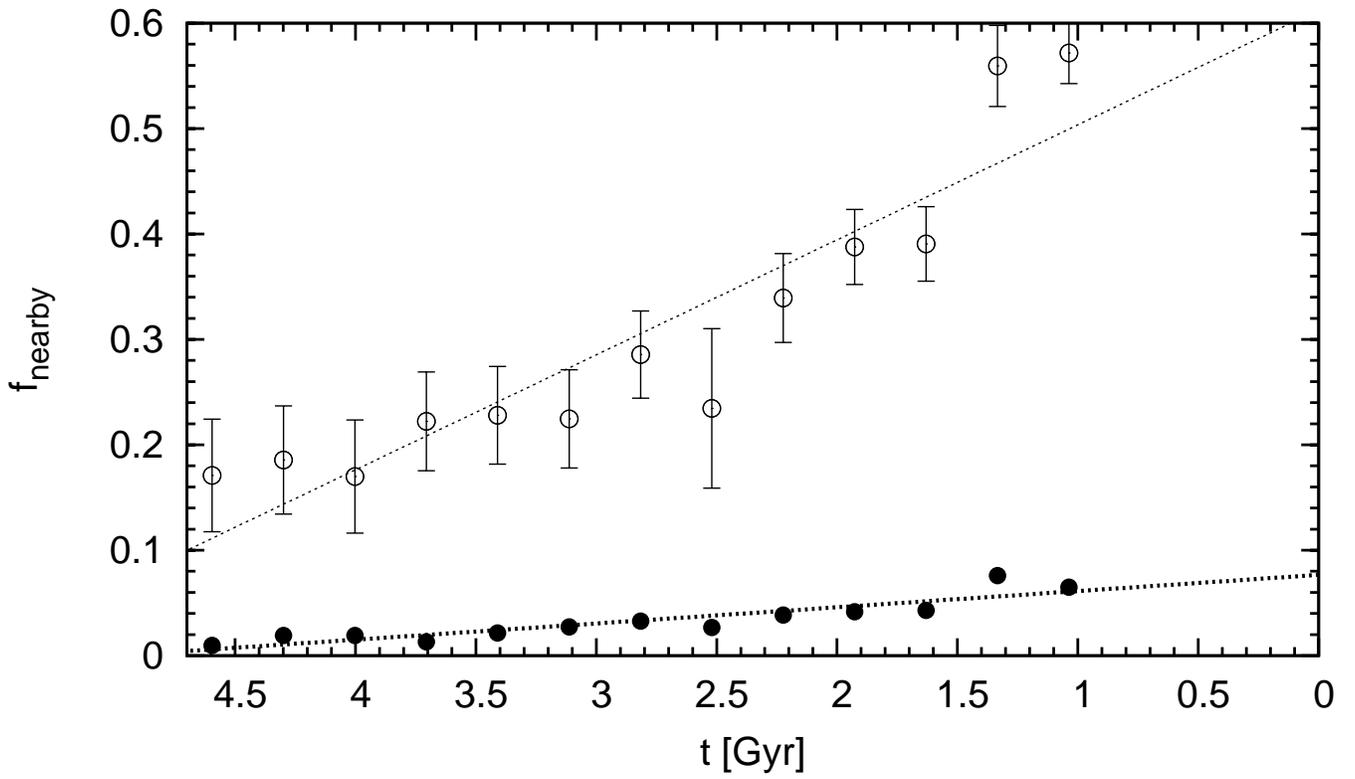,width=\columnwidth}
\caption[]{ 
   The fraction of siblings of the sun which are still in the
   neighborhood as a function of how long ago the parental cluster
   dissolved. The stars within 100\,pc (bullets) and those within 1kpc
   (circles) are fitted by a straight line (dotted curves).  The
   vertical error bars for the 1kpc calculations represent the one
   standard deviation from the mean and is based on calculating the
   trajectories of 2048 stars which were born in a virialized Plummer
   sphere with $M \simeq 920$\,\Msun\, and $R = 1$\,pc in the Galactic
   potential.  The fraction of nearby siblings depends only weakly on
   the uncertainty in the current position and velocity of the sun.  }
\label{Fig:NearbyFraction}
\end{figure}

\begin{figure}
\psfig{figure=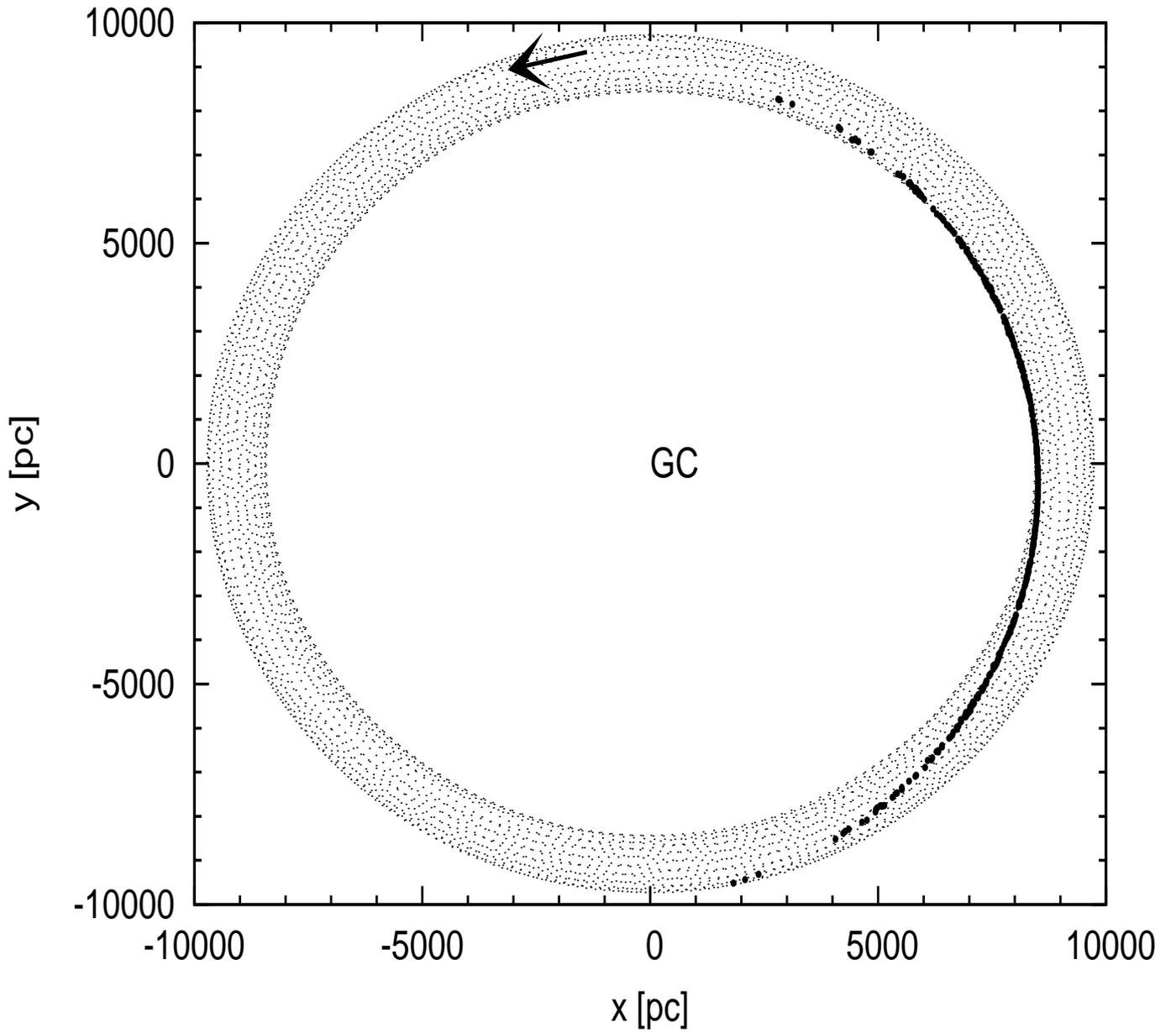,width=\columnwidth,height=0.9\columnwidth}
\caption[]{ 
Top view of the Galaxy. The orbit of the sun for the last 4.6\,Gyr is
presented with the thin dotted curve, starting at the arrow. The sun
is currently located at $x=8.5$kpc with $y=0$. A total of 1000 stars
from a star cluster with the parameters used in
Fig.\,\ref{Fig:NearbyFraction} were followed together with the sun and
evolved through time for 4.6\,Gyr, their final positions are plotted
as bullets.  }
\label{Fig:TopView}
\end{figure}

We persue by constructing model star clusters that mimic the one in
which the sun was born. These star clusters are assumed to dissolve
along the trajectory of the sun in the Galaxy, shedding their stars in
radial orbits with the escape speed of the cluster.  In that way we
are able to calculate the fraction of the stars from that particular
part of the orbit that today are still in the vicinity of the sun.  In
Fig.\,\ref{Fig:NearbyFraction} we present the fraction of stars that
can be found in the solar neighborhood as a function of the moment
that the star escaped from the parent cluster.

\placefigure{Fig:NearbyFraction}

The fraction of brothers and sisters of the sun that are still within
100\,pc of our current location in the Galaxy is about 1\% if the
proto-star cluster dissolved shortly after the formation of the sun,
and the fraction increases to $\sim 8$\% if the cluster survived
longer (see Fig.\,\ref{Fig:NearbyFraction}).  The fraction of siblings
within a kpc is at least $\sim 10$\,\%. The majority is located around
the past and future orbit of the sun in the Galactic disk and within
about 100\,pc of the Galactic plane. This is a consequence of the
relatively low velocity dispersion of the cluster compared to the
orbital motion around the Galactic center.

\begin{figure}
\psfig{figure=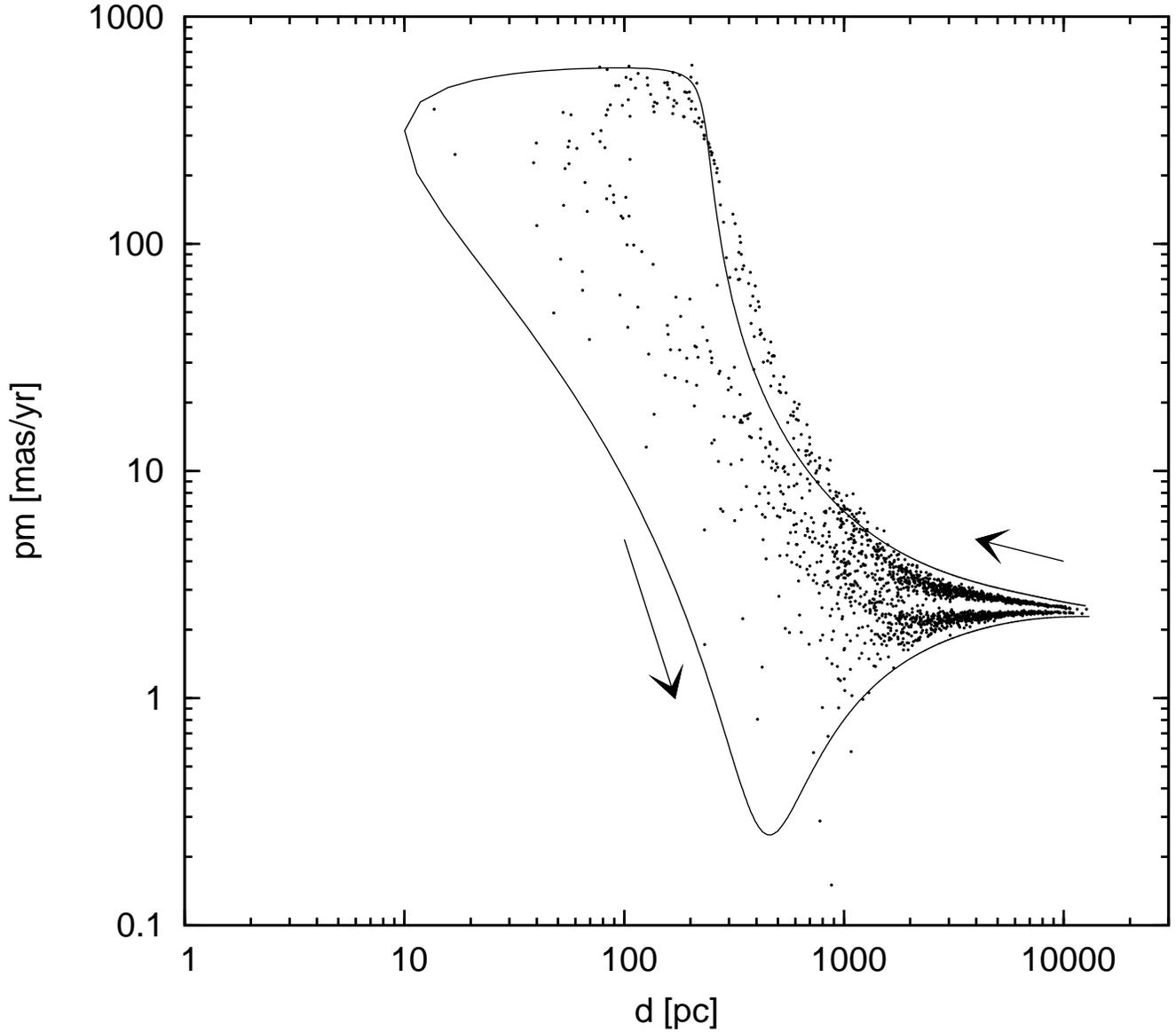,width=\columnwidth}
\caption[]{ 
The characteristic pattern of the proper motion and distance of the
stars in the solar neighborhood that once belonged to the same star
cluster. In this case we assumed that the star cluster dissolved
quickly upon the formation and pollution of the sun, $\sim 4.6$\,Gyr
ago.  The curve overplotted with the data points represent solar
orbit with a slight offset of 10\,pc along the x-axis and with a
10km/s higher velocity in the x-direction. The two arrows indicate how
the distance and proper motion change with time for stars that
approach the solar position (right arrow) or are ahead of the sun
(left arrow).}
\label{Fig:ProperMotion}
\end{figure}

Today, we should still be able to recognize the siblings along the
orbital trajectory of the solar system in the Galaxy.  If the sun
happens to be on a different orbit this will reflect in the orbits of
its siblings.  The distribution of proper motions and distances of
these stars are presented in Fig.\,\ref{Fig:ProperMotion} where we
show the result from a cluster of 2048 stars with a 1\,pc virial
radius that dissolved 4.6\,Gyr ago. The proper motions and distances
of these stars change in a characteristic way, indicated by the solid
curve and in the direction of the two arrows.

\placefigure{Fig:ProperMotion}

The majority of our lost siblings are easily identified by the Gaia
astrometric satellite or by ground based searches.  We are still
surrounded by them, even though they hide among millions of other
ordinary looking stars. What enables us to recognize the other
siblings are their orbital characteristics, which should be comparable
to the sun.  In Fig.\,\ref{Fig:TopView} we present the distribution of
stars that once belonged to the proto-solar cluster; they are
currently observable along the orbit of the sun in the Galaxy.  The
best place to look for the lost siblings is along the trajectory of
the sun in the plane of the Galaxy, in leading and trailing orbits
around the Galactic center.  Identifying those stars will provide
stringent limits on the sun's orbit around the Galactic center and
gives us a unique window to study the conditions of the star cluster
in which the sun was born.

\placefigure{Fig:TopView}



\section*{Acknowledgments}

This work was supported by the Netherlands Research School for
Astronomy (NOVA). I am grateful to Daan Porru for discussions and to
Douglas Heggie for comments on the manuscript.  This research was
supported in part by the National Science Foundation under Grant
No. PHY05-51164 and I thank the KITP in Santa Barbara for their
hospitality.


\end{document}